\documentclass[lettersize,journal]{IEEEtran}
\usepackage{amsmath,amsfonts}
\usepackage{algorithm}
\usepackage{algorithmic}
\usepackage{array}
\usepackage[caption=false,font=normalsize,labelfont=sf,textfont=sf]{subfig}
\usepackage{textcomp}
\usepackage{stfloats}
\usepackage{url}
\usepackage{verbatim}
\usepackage{amssymb}
\usepackage{graphicx}
\usepackage{balance}
\usepackage{cite}
\usepackage{xcolor}
\def\BibTeX{{\rm B\kern-.05em{\sc i\kern-.025em b}\kern-.08em
		T\kern-.1667em\lower.7ex\hbox{E}\kern-.125emX}}
\begin{document}
	\title{Exploiting Polarization Domain of the IOS for Enhanced Full-Dimensional Transmission}
	\author{Weiqiao Zhu, Zizhou Zheng, Yang Yang, Huan Huang, 
		
		Weijun Hao, Xiaofei Jia, and Hongliang Zhang
		\thanks{
			W. Zhu, Y. Yang, W. Hao, and X. Jia are with the Institute of Computing Technologies, China Academy of Railway Sciences Corporation Limited, Beijing 100081, China. (e-mail: zhuweiqiao@rails.cn; yangyang@rails.cn; haoweijun@rails.cn; carsrails@gmail.com).
			
			Z. Zheng is with the School of Electronics, Peking University, Beijing 100871, China, and the School of Information Science and Engineering, Southeast University, Nanjing, China. (e-mail: 213211229@seu.edu.cn).
			
			H. Huang is with the School of Electronic and Information Engineering, Soochow University, Suzhou, Jiangsu 215006, China. (hhuang1799@gmail.com).
			
			H. Zhang is with the School of Electronics, Peking University, Beijing 100871, China. (e-mail: hongliang.zhang92@gmail.com).

			\emph{Corresponding authors: Hongliang Zhang}
	}	
}
\markboth{}%
{How to Use the IEEEtran \LaTeX \ Templates}

\maketitle

\begin{abstract}
	Intelligent omni-surface (IOS), capable of providing service coverage to mobile users (MUs) in a reflective and refractive manner, has recently attracted
	widespread attention. However, the performance of power-domain IOS-assisted systems is limited by the intimate coupling between the refraction and reflection behavior of IOS elements.
	In this paper, we introduce the concept of dual-polarized IOS-assisted communication to overcome this challenge. By employing the polarization domain in the design of IOS, full independent refraction and reflection modes can be delivered. We consider a downlink dual-polarized IOS-aided system while also accounting for the leakage between different polarizations.
	To maximize the sum rate, we formulated a joint base station (BS) digital and IOS analog beamforming problem and proposed an iterative algorithm to solve the non-convex program. Simulation results validate that dual-polarized IOS significantly enhances the performance than that of the power-domain one.
\end{abstract}


\section{Introduction}
\IEEEPARstart{R}{ecently}, Intelligent omni-surface (IOS)-assisted wireless communication has attracted great attentions since it enables cost-effective and energy-efficient high data rate communication for future sixth-generation (6G) communication systems \cite{9690478}. It can provide service coverage to users on both sides of the surface and overcomes the limitation of widely studied reconfigurable intelligent surfaces (RISs), which only serve users on one side of the surface \cite{9696209,10319318}.

There has been extensive research on IOSs, demonstrating its potential to significantly enhance the service coverage of a base station (BS) compared to traditional RISs. 
In \cite {9491943}, multiple indoor users obtain omni-directionally services from a small base station with the aid of an IOS. In \cite{9889095}, the authors consider joint digital and analog beamforming at the BS and IOS, respectively. In these studies, the power ratio of the reflected and refracted signals is dependent on the structure of
the IOS elements and cannot be altered once designed \cite {9365009, 9086766}. Therefore, when the ratio of users on both sides is unbalanced with the power ratio, the performance of the power-domain IOS-assisted communication system deteriorates. 

To address this issue and realize the full potential of  IOS-aided systems, the concept of dual-polarized IOS is proposed \cite {10328189}. By utilizing the polarization domain, fully independent refraction and reflection modes can be delivered. The polarized IOS element can treat vertical and horizontal polarizations with polarization-independent electromagnetic (EM) responses.  Then, the power ratio of the refraction and reflection can be equivalently altered by allocating power to different polarized antennas at the BS.

Inspired by this, in this paper, we consider a downlink dual-polarized IOS-assisted system, while also accounting for the leakage between different polarizations. This polarization leakage has a significant impact on the system's performance, and a thorough analysis of the effects it causes is necessary. The main contributions are summarized as follows:
\begin{itemize}
	\item We propose a multi-user dual-polarized IOS-assisted downlink communication system. The physical characteristics of the IOS and	its polarization leakage are introduced and discussed.
	\item Based on this, we formulate a joint BS digital and IOS analog beamforming problem and aim to maximize the sum rate of the users on both reflective and refractive sides of the dual-polarized IOS.
	\item We further analyze the impact of the polarization leakage and user ratio on both sides of the IOS on the performance of the dual-polarized IOS-assisted system and compare the system with the power-domain IOS-aided one in the simulation. 
\end{itemize}

The rest of this paper is organized as follows. In Section II, we introduce the system model for dual-polarized IOS-assisted
communications and formulate the jointly digital and analog beamforming problem. In Section III, the original problem is reformulated as a more tractable form and divided into several sub-problems. Then, an iterative algorithm is proposed for addressing this problem. The theoretical analysis is elaborated in Section IV. Numerical results in Section V validate the performance enhancement of the dual-polarized IOS. Finally,conclusions are drawn in Section VI.
\begin{figure}[t]
	\centering
	\includegraphics[width=3in]{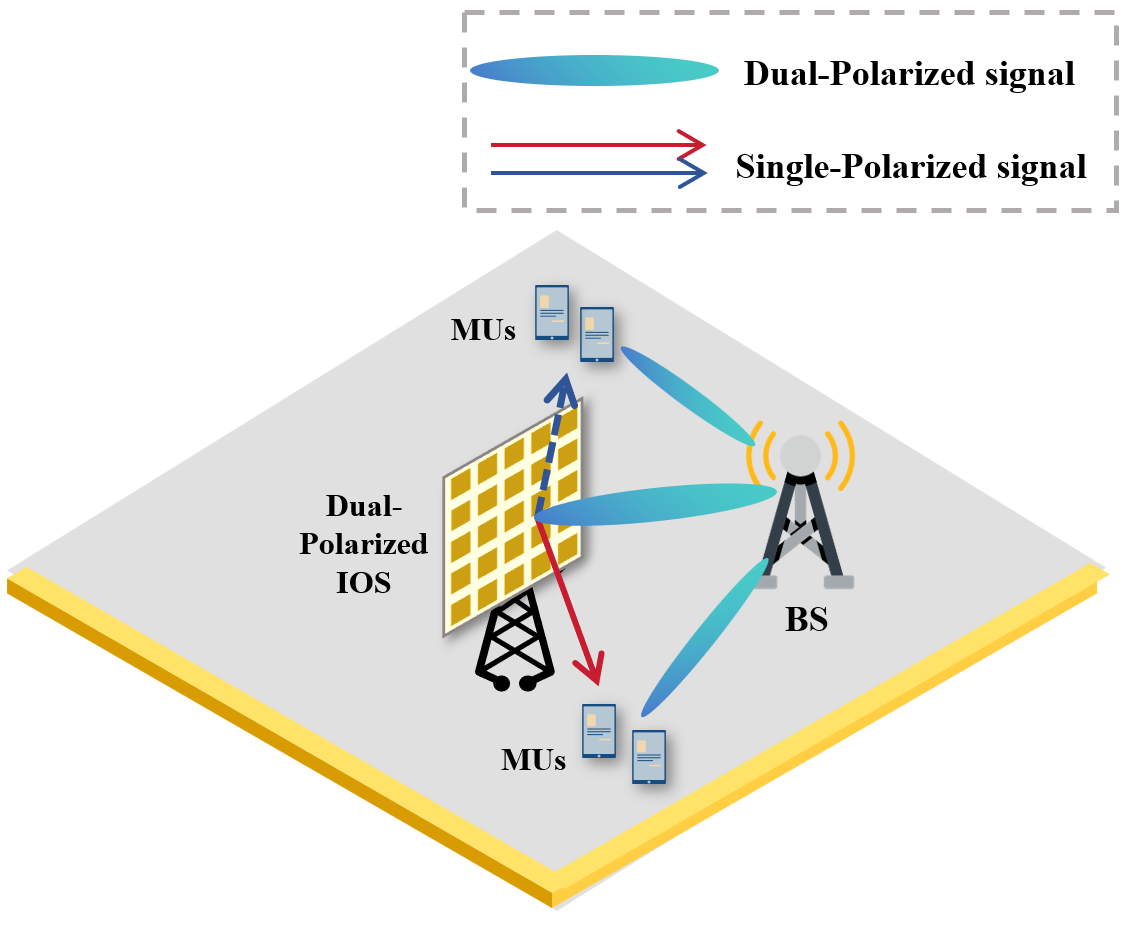}
	\caption{System model for a dual-polarized IOS-aided downlink system.}
	\label{fig_1}
\end{figure}
\section{System Model And Problem Formulation}
\subsection{Scenario Description}
\textcolor{red}{
As shown in Fig. 1, we consider a multi-user network where a BS equipped with \(N_t\) dual-polarized antennas communicates with \(2K\) mobile users (MUs). Each MU is equipped with a reconfigurable antenna that can select one of the two polarizations for signal reception \cite{10328189}. A dual-polarized IOS comprising \(M\) polarized elements is deployed between the BS and the MUs to reflect or refract transmitted signals, with the capability to select one polarization for reflection and the other for refraction.}

\textcolor{red}{
The set of all MUs is partitioned into two subsets based on their relative positions to the IOS. The subset \({\mathcal N}_{r}\) contains \(K_r\) MUs that receive signals reflected by the IOS, while the subset \({\mathcal N}_{t}\) contains \(K_t\) MUs that receive signals refracted by the IOS. The total number of users satisfies ${\mathcal N}_{r}\cup {\mathcal N}_{t}=2K$.}

\subsection{System Model}
The signals transmitted from the BS are expressed as 
\begin{align}
	{\bf x}=\left[\begin{array}{cc}
		\mathbf{W}_r, \mathbf{W}_t
	\end{array}\right]{\bf s}=\left[\begin{array}{l}
		{\mathbf{x}}_{v}\\
		{\mathbf{x}}_{h}
	\end{array}\right] \in \mathbb{C}^{2 N_{{t}} \times 1}, \label{1}
\end{align}
where ${\bf s} \sim \mathcal {CN}(0, {\bf I}) \in \mathbb C^{2 K \times 1}$ denotes different data streams for the MUs, ${\bf W}=[{\bf W}_{r}, {\bf W}_{t}]$ denotes the digital beamforming matrix, and ${\bf x}_{v}$, ${\bf x}_{h}$ represents the signal transmitted from the vertically and horizontally polarized antennas, respectively.

\subsubsection{Dual-polarized IOS}The EM response $g_m$ of the $m^{th}$ IOS element takes different values depending on the signal's polarization state and is given by
\begin{align}
	g_{m}^{\mathcal{X}}=\left\{\begin{array}{ll}
		g_{m}^{{vv}}, & \text { vertically polarized, } \\
		g_{m}^{{hh}}, & \text { horizontally polarized. }
	\end{array}\right.
\end{align}
where we define $\mathcal{X}\in\{vv,hh\}$ as an indicator. 

Furthermore, $g_m$ is an angle-dependent value and can be modeled as 
\begin{align}
	g_{m}^{\mathcal{X}}=\sqrt{G_{m} F_{m}\left(\theta_{i, m}^{\mathcal{X}}, \phi_{i, m}^{\mathcal{X}}\right) F_{m}\left(\theta_{r, m}^{\mathcal{X}}, \phi_{r, m}^{\mathcal{X}}\right) S_{m}} e^{j \psi^{\mathcal X}_{m}},
\end{align}
where $G_m$ is the power gain, $S_m$ is the surface area of the IOS element and $F_m$ is the normalized radiation intensity of the IOS element in different directions. $(\theta_{i, m}^{\mathcal{X}}, \phi_{i, m}^{\mathcal{X}})$ is the incident angle from the transmitting antenna to the $m^{th}$ element, $(\theta_{r, m}^{\mathcal{X}}, \phi_{r, m}^{\mathcal{X}})$ is the departure angle from the $m^{th}$ element to the receiving antenna, and $\psi^{\mathcal X}_{m}$ are the phase shift adjustments that are imposed on the signal. 

We assume that the dual-polarized IOS has $N$-bits quantized phase shifts. Then, the available phase shift set is denoted by ${\mathcal S}_a=\{0, ..., \frac{2\pi}{2^N}(2^N-1)\}$ and the phase shift of the $m^{th}$ element $\psi_{m} \in {\mathcal S}_a$.

It is noted that for the dual-polarized IOS, the phase adjustments of the refracted and reflected link can be independently controlled due to the use of the polarization domain of the dual-polarized IOS, while for power-domain IOS, the refraction and reflection coefficients are intimately coupled due to the EM properties of the surface \cite {9774942}.

\subsubsection{Channel Model} In an ideal situation, the EM waves in orthogonal polarizations exhibit nearly independent propagation characteristics. However, the scatterers in the environment can cause a cross-polarization effect that alters the polarization state of the EM waves and then produces polarization leakage, which can be measured by cross-polarization discrimination (XPD).

Take the channel between the BS and the IOS as an example, which is denoted by
\begin{align}\label {ch1}
	{\mathbf{H}}_{\mathrm{BI}}=\left[\begin{array}{ll}
		{\mathbf{H}}_{\mathrm{BI}}^{{vv}} & {\mathbf{H}}_{\mathrm{BI}}^{{vh}} \\
		{\mathbf{H}}_{\mathrm{BI}}^{{hv}} & {\mathbf{H}}_{\mathrm{BI}}^{{hh}}
	\end{array}\right] \in \mathbb{C}^{2 M \times 2 N_{{t}}},
\end{align}
where ${\mathbf{H}}_{\mathrm{BI}}^{{vv}}$ and $	{\mathbf{H}}_{\rm{BI}}^{{hh}}$ are the co-polarized components and ${\mathbf{H}}_{\rm{BI}}^{{vh}}$ and $	{\mathbf{H}}_{\rm{BI}}^{{hv}}$ are the cross-polarized components. The XPD of ${\mathbf H}_{\rm BI}$ can be expressed as
\begin{align}\label{xpd}
	\text { XPD }=\frac{{\mathbb E}\{\|{\mathbf{H}}_{\mathrm{BI}}^{{vv}}\|^2_F\}}{{\mathbb E}\{\|{\mathbf{H}}_{\rm{BI}}^{{hv}}\|^2_F\}}=\frac{{\mathbb E}\{\|{\mathbf{H}}_{\mathrm{BI}}^{{hh}}\|^2_F\}}{{\mathbb E}\{\|{\mathbf{H}}_{\rm{BI}}^{{vh}}\|^2_F\}}=\frac{1-\beta_{\rm BI}}{\beta_{\rm BI}},
\end{align}
where $\beta_{\rm BI}$ is the XPD factor.

The channel ${\bf h}_{{\rm IU},k}$ between the IOS and the $k^{th}$ user is defined as
\begin{align}\label{ch2}
	{\bf h}_{{\rm IU},k}=
	\begin{cases} 
		[{\bf h}^{vv}_{{\rm IU},k}, {\bf h}^{vh}_{{\rm IU},k}], k\in {\mathcal N}_{r} \\
		[{\bf h}^{hv}_{{\rm IU},k}, {\bf h}^{hh}_{{\rm IU},k}], k\in {\mathcal N}_{t},
	\end{cases}
\end{align}
where the XPD factor is defined as $\beta_{{\rm IU},k}$. 
The direct path ${\bf h}_{{\rm BU},k}$ between the BS and the $k^{th}$ MU is modeled likewise.
\subsection{Problem formulation}
The received signal at the $k^{th}$ MU is expressed as 
\begin{align}\label{receive}
	y_k={\bf h}_{k}{\bf x}+n_k,
\end{align}
where $n_k$ is the additive white Gaussian noise (AWGN) at the $k^{th}$ user whose mean is zero and variance is $\sigma^2$.
The overall channel between the BS and the $k^{th}$ user is written as
\begin{align}\label{ch2}
	{\bf h}_{k}=
	{\bf h}_{{\rm BU},k}+{\bf h}_{{\rm IU},k}{\bf G}{\bf H}_{\rm BI},
\end{align}
where the IOS coefficient matrix $\bf G$ is denoted by ${\rm diag}\{g^{vv}_1,...,g^{vv}_m, g^{hh}_1,...,g^{hh}_m\}$. 

Assume that the channel state information (CSI) is known at the BS. Based on the received signal in (\ref {receive}),  the signal-to-interference-plus-noise ratio (SINR) of the $k_r \in {\mathcal N}_{r}$ MU in the reflective side can be formulated as
\begin{align}\label{sumrate}
	\gamma_{k_r} =\frac{|{\bf h}_{r,k_r}{\bf W}_{r}^{(k_r)}|^2}{\sum\limits_{k_r'\neq k_r }^{}|{\bf h}_{r,k_r}{\bf W}_{r}^{(k_r')}|^2+\sum\limits_{i\in {\mathcal N}_{t}}|{\bf h}_{r,k_r}{\bf W}_{t}^{(i)}|^2+\sigma^2} ,
\end{align}
where ${\bf W}^{(k)}$ denotes the $k^{th}$ column of the beamformer $\bf W$, $\sum_{k_r'\neq k_r }^{}|{\bf h}_{r,k_r}{\bf W}_{r}^{(k_r')}|^2$ denotes the interference of MUs with same polarization, and $\sum_{i\in {\mathcal N}_{t}}|{\bf h}_{r,k_r}{\bf W}_{t}^{(i)}|^2$ represents the polarization leakage of other MUs with different polarization. The SINRs of the MUs on the refractive side of the IOS can be similarly written. 

The optimization problem can be formulated as
\begin{subequations}\label{w1}
	\begin{align} 
		&\max_{\substack{ {\bf W}_{r}, {\bf W}_{t} \\ \{\psi^{vv}_{m}\}, \{\psi^{hh}_{m}\} }}   \sum_{{k_r \in \mathcal N}_r}^{} {\log_2(1+\gamma_{k_r})}  + \sum_{k_t \in {\mathcal N}_t}^{} {\log_2(1+\gamma_{k_t})} \label{ori}\\
		&\qquad \quad \text{s.t.}\quad{\rm Tr}({\bf W}_{r}{\bf W}_{r}^{\rm H})+{\rm Tr}({\bf W}_{t}{\bf W}_{t}^{\rm H}) \le P_{\rm BS}, \label{o1}\\
		&\qquad \quad \qquad\psi^{\mathcal X}_m \in {\mathcal S}_a \label{o3},
	\end{align}
\end{subequations}
where $P_{\rm BS}$ denotes the maximum transmit power allowed by the BS.

\section{Dual-Polarized IOS-based Beamforming Algorithm}
In this section, we first transform (\ref {w1}) into a more tractable one, which decouples the optimization of the digital and analog beamforming at the BS and the IOS, respectively. Then, the original problem can be separated into several sub-problems.
\subsection{Reformulate of the original problem}
The minimum mean-square error (MMSE) algorithm is employed for addressing the non-convex (\ref {ori}). By introducing the auxiliary variables ${\bf f}= [f_{r,1}, ...,f_{r,K_r}, f_{t,1}..., f_{t,K_t}]^{\rm T}$, and ${\bf u}=[u_{r,1},..., u_{r,K_r}, u_{t,1}..., u_{t,K_t}]^{\rm T}$, Problem (\ref {w1}) can be reformulated as 
\begin{subequations}\label{ne}
	\begin{align} 
		&\max_{\substack{ {\bf W}_{r}, {\bf W}_{t} \\ {\bf G}, {\bf f}, {\bf u}}} \quad \sum_{{k_r \in \mathcal N}_r}^{} \left({\rm log}_2f_{r,k_r}-f_{r,k_r}e_{r,k_r}\right)  \nonumber
		\\ &\qquad \qquad +\sum_{k_t \in {\mathcal N}_t}^{} \left({\rm log}_2f_{t,k_t}-f_{t,k_t}e_{t,k_t}\right)  \label{mseobj} \\
		&\quad \text{s.t.}  \quad (\rm \ref {o1}), (\rm \ref {o3}), 
	\end{align}
\end{subequations}
where the MSE of the ${k_r}^{th}$ MU receiving the reflected signal is given by 
\begin{align}
	\operatorname{e}_{{r}, k_{{r}}}= & \mathbb{E}\left\{\left({u}_{{r}, k_{{r}}}^{*} y_{{r}, k_{{r}}}-s_{{r}, k_{{r}}}\right)\left(u_{{r}, k_{{r}}}^{*} y_{{r}, k_{{r}}}-s_{{r}, k_{{r}}}\right)^{*}\right\} \label{mse} \nonumber \\
	= & \sum_{i \in \mathcal{N}_{{r}}}\left|u_{{r}, k_{{r}}}^{*}{\bf h}_{r,k_r} \mathbf{W}^{(i)}_{{r}}\right|^{2} +\sum_{j \in \mathcal{N}_{{t}}}\left|u_{{r}, k_{{r}}}^{*}{\bf h}_{r,k_r} \mathbf{W}^{(j)}_{{t}}\right|^{2}   \nonumber \\
	& -2 {\rm{Re}}\left\{u_{{r}, k_{{r}}}^{*}{\bf h}_{r,k_r} \mathbf{W}^{(k_r)}_{{r}}\right\} +\left|u_{{r}, k_{{r}}}\right|^{2} \sigma^{2}+1. 
\end{align}

By taking the first derivative of (\ref {mseobj}) equal to zero, we obtain the optimal solutions of $f_{r,k_r}$ and $u_{r,k_r}$ as
\begin{align}
	&u_{r,k_r}^{\star}=\frac{{\bf h}_{r,k_r}{\bf W}_{r}^{(k_r)}}{\|{\bf h}_{r,k_r}{[{\bf W}_{r},{\bf W}_{t}]}\|^2+\sigma^2},\label{u} \\
	&f_{r,k_r}^{\star}= \frac{1}{  1-u_{r,k_r}^{op} {\bf W}_{r}^{(k_r)\rm H} {\bf h}_{r,k_r}^{\rm H} }.\label{f}
\end{align}

In the following, we focus on the optimization of the digital and analog beamformer ${\bf W}_r, {\bf W}_t$ and $\bf G$.

\subsection{Digital beamforming optimization at the BS}
With fixed $\bf f$, $\bf u$, and $\bf G$, Problem (\ref {ne}) is rewritten as
\begin{subequations}\label{wop}
	\begin{align} 
		&\min_{{\bf W}_r, {\bf W}_t} \sum_{{k_r \in \mathcal N}_r} f_{r,k_r}(\|u^*_{r,k_r}{\bf h}_{r,k_r}{\bf W}\|^2-2{\rm {Re}}\{u_{r,k_r}^*{\bf h}_{r,k_r}{\bf W}_{r}^{(k_r)}\})  \nonumber 
		\\ &+\sum_{k_t \in {\mathcal N}_t}^{} f_{t,k_t}(\|u^*_{t,k_t}{\bf h}_{t,k_t}{\bf W}\|^2-2{{\rm {Re}}\{u_{t,k_t}^*{\bf h}_{t,k_t}{\bf W}_{t}^{(k_t)}\}})   \\
		&\quad \text{s.t.}  \quad {\rm Tr}({\bf W}_{r}{\bf W}_{r}^{\rm H})+{\rm Tr}({\bf W}_{t}{\bf W}_{t}^{\rm H}) \le P_{\rm BS}.
	\end{align}
\end{subequations}
The Lagrangian function of Problem (\ref {wop}) is written as
\begin{align}
	&\mathcal L({\bf W}_r, {\bf W}_t, \lambda)=
	\lambda ( \sum\limits_{i \in {\mathcal N}_r}{\bf W}_r^{(i)\rm H}{\bf W}_r^{(i)} +\sum\limits_{j \in {\mathcal N}_t}{\bf W}_t^{(j)\rm H}{\bf W}_t^{(j)}  )\nonumber
	\\&-\sum_{i\in {\mathcal N}_{r}}2f_{r,i}{{\rm {Re}}\{u_{r,i}^{*}{\bf h}_{r,i}{\bf W}_{r}^{(i)}\}}-\sum_{j\in {\mathcal N}_{t}}2f_{t,j}{{\rm {Re}}\{u_{t,j}^{*}{\bf h}_{t,j}{\bf W}_{t}^{(j)}\}}
	\nonumber  
	\\&+\sum_{i\in {\mathcal N}_r}{\bf W}_{r}^{(i)\rm H}{\bf M}{\bf W}_{r}^{(i)}+\sum_{j\in {\mathcal N}_t}{\bf W}_{t}^{(j)\rm H}{\bf M}{\bf W}_{t}^{(j)}-\lambda P_{\rm {BS}},
\end{align}
where ${\bf M}=\sum_{} f_{r,i}|u_{r,i}|^2{\bf h}_{r,i}^{\rm H}{\bf h}_{r,i}+\sum_{} f_{t,j}|u_{t,j}|^2{\bf h}_{t,j}^{\rm H}{\bf h}_{t,j}$, and $\lambda \ge 0$ is the Lagrangian multiplier with the constraint (\ref {o1}).

By setting the first-order derivative of $\mathcal L({\bf W}_{r}, {\bf W}_{t}, \lambda)$ w.r.t. ${\bf W}_r^{(k_r)}, {\bf W}_{t}^{(k_t)}$ to zero, the optimal solution is given by
\begin{align}
	&\mathbf{W}_{r}^{(k_r)\star}\left(\lambda\right)=u_{r,k_r}f_{r,k_r}\left({\bf M}+\lambda \mathbf{I}\right)^{\dagger}{\bf h}_{r,k_r}^{\rm H}, \label {wr}\\
	&\mathbf{W}_{t}^{(k_t)\star}\left(\lambda\right)=u_{t,k_t}f_{t,k_t}\left({\bf M}+\lambda \mathbf{I}\right)^{\dagger}{\bf h}_{t,k_t}^{\rm H}, \label{wt}
\end{align}
where  (·)$^\dagger$ denotes the matrix pseudo-inverse. The value of $\lambda$ should satisfy the complementary slackness condition $	\lambda(\sum_{i \in {\mathcal N}_r}{\bf W}_{r}^{(i)\rm H}{\bf W}_{r}^{(i)}+\sum_{j \in {\mathcal N}_t}{\bf W}_{t}^{(j)\rm H}{\bf W}_{t}^{(j)}-P_{\rm BS})=0$.
The value of $\lambda$ should be chosen to ensure the following equation:
\begin{align}\label{21}
	{\mathcal F}(\lambda) \triangleq 	\sum_{i=1}^{2N_t}\frac{\left[\sum_{k=1}^{2K}|u_kf_k|^2{\bf Q}^{\rm H}{\bf h}_{k}^{\rm H}{\bf h}_k{\bf Q}\right]_{i,i}}{([{\bf \Lambda}]_{i,i}+\lambda)^2}=P_{\rm BS}.
\end{align}
where [·]$_{i,i}$ denotes the $i$th diagonal element of the matrix. 
The unitary matrix $\bf Q$ and the diagonal matrix $\bf \Lambda$ are derived from the singular value decomposition (SVD) of the matrix ${\bf M}={\bf Q}{\bf \Lambda}{\bf Q}^{\rm H}$, where ${\bf Q}{\bf Q}^{\rm H}={\bf Q}^{\rm H}{\bf Q}={\bf I}_{2N_t}$.

Readily, the total power consumption ${\mathcal F}(\lambda)$ is a decreasing function of $\lambda$ and ${\mathcal F}(\infty)=0$. Thus, the bisection search method can be employed to find $\lambda$, and the upper bound of $\lambda$ is given by $\lambda \le \sqrt{\frac{\sum_{i=1}^{2N_t}\left[\sum_{k=1}^{2K}|u_kf_k|^2{\bf Q}^{\rm H}{\bf h}_{k}^{\rm H}{\bf h}_k{\bf Q}\right]_{i, i}}{P_{\rm BS}} }=\lambda^{ub}.$
\subsection{Analog beamforming optimization at the IOS}
Given $\bf f$, $\bf u$, ${\bf W}_{r}$, and ${\bf W}_{t}$, the discrete IOS analog beamforming subproblem can be written as
\begin{subequations}\label{gop}
	\begin{align} 
		&\min_{{\bf G}} \quad  {\rm Tr}({\bf G}^{\rm H}{\bf B}{\bf G}{\bf C})+ 2{\rm Re}\{{\rm Tr}(\bf GV)\} \label{objg}\\
		&\quad \text{s.t.}  \quad (\rm \ref {o3}),
	\end{align}
\end{subequations}
where ${\bf B}= \sum_{k_r \in {\mathcal N}_r}f_{r,k_r}u_{r,k_r}u_{r,k_r}^{*}{\bf h}_{{\rm IU},k_r}^{\rm H}{\bf h}_{{\rm IU},k_r}+\sum_{k_t \in {\mathcal N}_t}f_{t,k_t}u_{t,k_t}u_{t,k_t}^*{\bf h}_{{\rm IU},k_t}^{\rm H}{\bf h}_{{\rm IU},k_t}$, ${\bf C}={\bf H}_{\rm BI}{\bf W}{\bf W}^{\rm H}{\bf H}_{\rm BI}^{\rm H}$, and ${\bf V}=\sum_{i\in {\mathcal N}_{r}}f_{r,i}u_{r,i}^*(u_{r,i}{\bf H}_{\rm BI}{\bf W}{\bf W}^{\rm H}{\bf h}_{{\rm BU},i}^{\rm H}{\bf h}_{{\rm IU},i}-{\bf H}_{\rm BI}{\bf W}_{r}^{(i)}{\bf h}_{{\rm IU},i})+ \sum_{j\in {\mathcal N}_{t}}f_{t,j}u_{t,j}^*(u_{t,j}{\bf H}_{\rm BI}{\bf W}{\bf W}^{\rm H}{\bf h}_{{\rm BU},j}^{\rm H}{\bf h}_{{\rm IU},j}-{\bf H}_{\rm BI}{\bf W}_{t}^{(j)}{\bf h}_{{\rm IU},j})$.

The objective function in (\ref {gop}) is a convex function of
the IOS coefficient matrix $\bf G$. By releasing the non-convex  (\ref {o3}), it can be solved by standard tools such as CVX. However, $\psi^{{\mathcal X}, opt}_m$ lies in the range of $l_m\Delta \psi$ and $(l_m+1)\Delta \psi$, where $l_m$ denotes an integer satisfying $0\le l_m\Delta \psi \le \frac{2\pi}{2^N}(2^N-1)$, and $\Delta \psi= \frac{2\pi}{2^N}$ represents the discrete phase shift step, which can be addressed by the BnB based algorithm \cite {9491943}.

\subsection {Overall Algorithm}
Based on the above analysis, we provide the detailed steps of the algorithm to solve Problem (\ref {w1}) in Algorithm 1. 
\begin{algorithm}[t]
	\caption{Joint Digital and Analog Beamforming Design for Sum-Rate Maximization Problem}
	\begin{algorithmic}[1]
		\STATE
		\text{Initialize} \(\mathbf{W}_{r}, \mathbf{W}_{t}, \{\psi^{vv}_{m}\}, \{\psi^{hh}_{m}\}  \).
		\STATE 
		\textbf{While} \text{no convergence of the objective function (\ref {mseobj}) }
		\STATE
		\quad \text{Calculate} \(\bf f\) \text{by} \((\ref {f})\) and \(\bf u\) \text{by} \((\ref {u})\);
		\STATE
		\quad \text{Calculate} \(\bf G\) by solving Problem (\ref {gop}) relaxing (\ref {o3});
		\STATE 
		\quad \text{Calculate the optimal} $\bf G$ with discrete phase shifts by adopting BnB algorithm;
		\STATE
		\quad \text{Calculate} \({\bf W}\) with the Lagrangian multiplier method;
		\STATE	
		\textbf{Until} \text{the objective function (\ref {mseobj}) converges.}
		
	\end{algorithmic}
	\label{alg2}
\end{algorithm}

\textcolor{red}{We denote the sum-rate at the $i^{th}$ iteration by $R({\bf W}^{i},{\bf G}^{i})$. At the $(i+1)^{th}$ iteration, the solution of the optimal IOS analog beamforming given by ${\bf W}^{i}$ yields a sum-rate $R({\bf W}^{i},{\bf G}^{i+1}) \ge R({\bf W}^{i},{\bf G}^{i})$. Similarly,  the solution of the optimal BS digital beamforming given by ${\bf G}^{i+1}$ yields  $R({\bf W}^{i+1},{\bf G}^{i+1}) \ge R({\bf W}^{i},{\bf G}^{i+1})$. Therefore, we obtain the following inequalities $R({\bf W}^{i+1},{\bf G}^{i+1}) \ge R({\bf W}^{i},{\bf G}^{i})$. In addition, due to the total transmit power constraint, the sum rate has an upper bound. Hence, Algorithm 1 is guaranteed to converge.}

\textcolor{red}{
The main complexity of Step 6 lies in calculating $\mathbf{W}^{(k)}$ in (\ref {wr})-(\ref {wt}), especially in the matrix inversion of ${\bf M} \in {\mathbb C}^{2N_t\times 2N_t}$. Therefore, the complexity is in the order of $\mathcal {O}(KN_t^3)$. The complexity of analog beamforming problem is the sum of the continuous and discrete phase shift subproblems. The continuous one lies in the interior point method and can be referred to in \cite{10319318}. Hence, it is with the complexity of ${\mathcal O}(M^{4.5})$, and the order of the BnB algorithm is, in the worst case, $\mathcal{O}({M^{4.5}2^{M}})$ \cite{9491943}. Thus, the total complexity is $\mathcal {O}({KN_t^3+M^{4.5}2^{M}})$.}

\section{Performance Analysis of the Dual-polarized IOS-assisted communication system}
\subsection{Analysis of the XPD}
The power allocated to the vertical and horizontal polarization is expressed as
\begin{align}
	&p_{v}={\rm Tr}(\frac{\sum_{k=1}^{K_r}|u_{r,k_r}f_{r,k_r}|^2{\bf h}_{r,k_r}^{\rm H}{\bf h}_{r,k_r}}{({\bf M}+\lambda {\bf I})^2}), \\
	&p_{h}={\rm Tr}(\frac{\sum_{k=1}^{K_t}|u_{t,k_t}f_{t,k_t}|^2{\bf h}_{t,k_t}^{\rm H}{\bf h}_{t,k_t}}{({\bf M}+\lambda {\bf I})^2}).
\end{align}
It can be noted that the factor $\lambda$ does not affect the power distribution between the two polarization directions. Polarization leakage caused by the channel is what affects the power distribution. Then, we analyze the impact of polarization leakage, i.e., XPD on the digital beamformer ${\bf W}=[{\bf W}_r, {\bf W}_t]$.

\textit{Proposition 1:} If the polarization leakage is zero, each MU's digital beamformer ${\bf W}^{(k)}$ will have half of its elements corresponding to the antennas in the opposite polarization direction equal to zero.

\textit{Proof:} When there is no polarization leakage, the channel's cross-polarization component does not exist. From the expressions in (\ref {wr}) and (\ref {wt}), it is evident that the conclusion holds.

\textit{Proposition 2:} As the cross-polarization factor ${\beta}_{\rm BI}$ given in (\ref {xpd}) increases from 0 to 1, the system performance first degrades and then becomes better.

\textit{Proof:} Note that when ${\beta}_{\rm BI}=0$, the polarization of the transmitted signal will not change when it propagates from the BS. Alternatively, ${\beta}_{\rm BI}=1$ corresponds to the case that the transmitted signal will be completely converted to the opposite polarization and can also be decomposed as two dependent subchannels. On the other hand, if $0 <\beta_{\rm BI} < 1$, a part of the power will be leaked to the other polarization, which causes cross-polarization interference and degrades the performance. 

Based on  \textit{Proposition 2}, we can obtain \textit{Remark 1}.

\textit{Remark 1:} High polarization leakage may provide extra gains for dual-polarized IOS-assisted systems.

\subsection{Analysis of the user ratio ${K}_{r}/{K}_{t}$}

The sum rate of power-domain IOS-assisted systems is impacted by ${K}_{r}/{K}_{t}$ since the elements have a fixed power ratio. When the power ratio is unmatched with the user ratio on both sides, the performance is poor. A dual-polarized IOS-assisted system can overcome this by power allocation for different polarized antennas at the BS. 

\textit{Remark 2:} The sum rate of dual-polarized IOS-assisted systems is not impacted by the user ratio ${K}_{r}/{K}_{t}$.
%
%
\begin{figure*}[h]
	\setlength{\abovecaptionskip}{-5pt}
	\setlength{\belowcaptionskip}{-10pt}
	\centering
	\begin{minipage}{0.32\linewidth}
		\vspace{3pt}
		\includegraphics[width=2.54in]{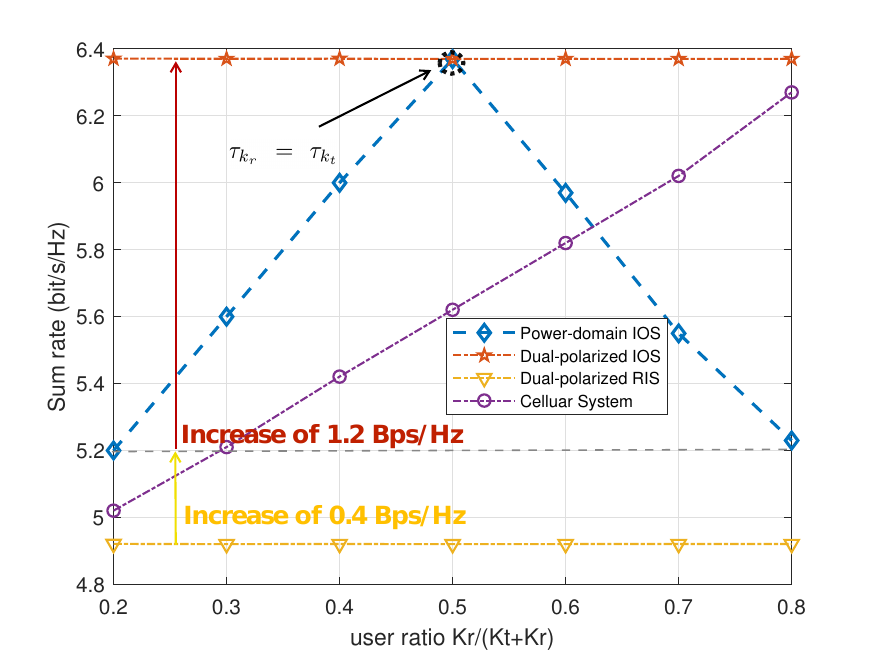}
		\caption{Sum rate versus $K_r/(K_t+K_r)$.}
		\label{fig_4}
	\end{minipage}
	\begin{minipage}{0.32\linewidth}
		\vspace{3pt}
		\includegraphics[width=2.54in]{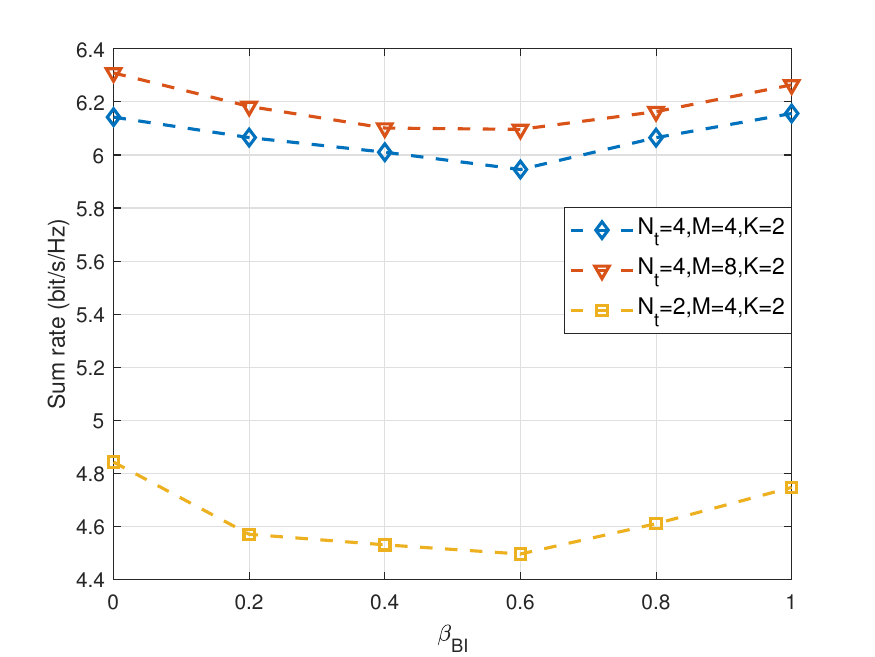}
		
		\caption{Sum rate versus XPD.}
		\label{fig_2}
	\end{minipage}
	\begin{minipage}{0.32\linewidth}
		\vspace{3pt}
		\includegraphics[width=2.54in]{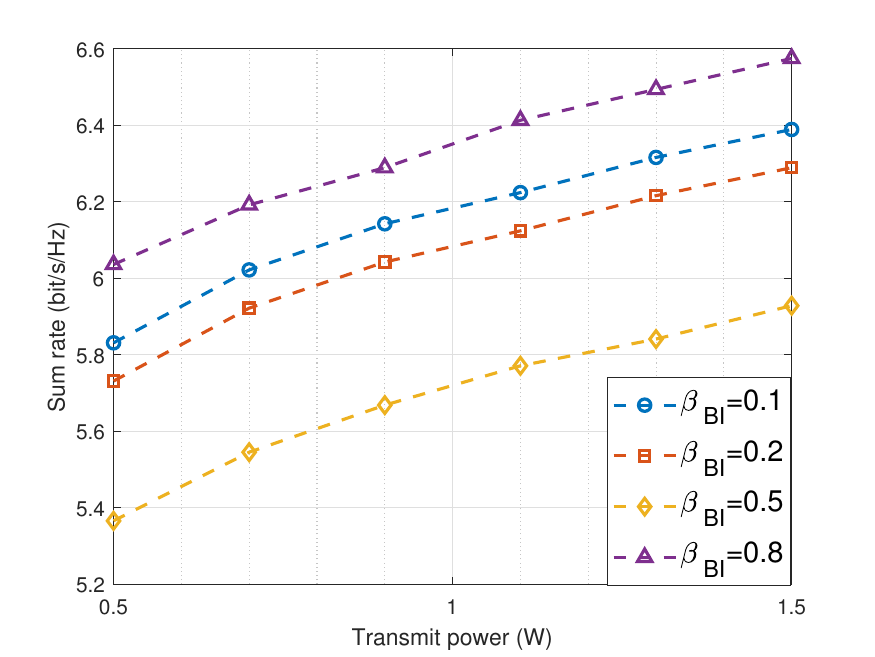}
		
		\caption{Sum rate versus transmit power.}
		\label{fig_3}
	\end{minipage}
	
\end{figure*}

\textit{Proposition 4:} When the number of users on both sides of the IOS is the same, symmetrically distributed, and the power allocation ratio is 1, the direct link leads to a performance gap between the power-domain IOS and the dual-polarized IOS.

\textit{Proof:} Considering a power-domain IOS with power-split ratio $\epsilon$, the sum rate of the MUs on both sides is expressed as 
\begin{align}\label{sr}
	\sum_{k=1}^{K_r+K_t}R_k= \sum_{{k_r \in \mathcal N}_r}^{} &{\log_2(1+\tau_{k_r}+\frac{\epsilon}{1+\epsilon}\chi_{k_r})} \nonumber \\ &+ \sum_{k_t \in {\mathcal N}_t}^{} {\log_2(1+\tau_{k_t}+\frac{1}{1+\epsilon}\chi_{k_t})},
\end{align}
where $\tau_{k}$ is the SINR of ${k}^{th}$ MU from direct link, $\chi_k$ is the SINR of $k^{th}$ MU from the IOS-aided link. By taking the first derivative of (\ref{sr}) with respect to $\epsilon$ , we obtain
\begin{align}
	\sum_{k=1}^{K_r+K_t}\frac{dR_k}{d\epsilon}= 
	 \frac{1}{\rm ln2}&\sum_{k_t \in {\mathcal N}_t} {\frac {\chi_{k_t}/(1+\epsilon)^2}{1+\tau_{k_t}+\frac{\epsilon}{1+\epsilon}\chi_{k_t}}} \nonumber \\
	&-\frac{1}{\rm ln2}\sum_{k_r \in {\mathcal N}_r} {\frac {\chi_{k_r}/(1+\epsilon)^2}{1+\tau_{k_r}+\frac{1}{1+\epsilon}\chi_{k_r}}}.
\end{align}
Here, we assume $\chi_{k_t}=\chi_{k_r}=\chi_{k}$. If $\tau_{k_r}=\tau_{k_t}$, we have $\epsilon^{opt}=1$ and there is no performance gap between the power-domain and dual-polarized IOS. Otherwise, $\epsilon^{opt} \neq 1$ and a performance gap exists.

\section{Simulation Results}
\textcolor{red}{In this section, we evaluate the performance of the considered dual-polarized IOS-assisted system based on the proposed algorithm and compare it with a power-domain IOS-aided system, a dual-polarized RIS-aided system and a cellular system. In the conventional cellular system, the MUs receive only the direct links from the BS without the assistance of an IOS. The details of the simulation parameters are as follows: number of BS antennas $N_t=4$, number of IOS elements $M=4$, number of MUs $K_r=K_t=2$, XPD factors $\beta_{\rm BI}=\beta_{\rm IU}=\beta_{\rm BU}=0.1$, the transmit power $P_{\rm BS}=1 {\rm W}$. The values of XPD are chosen according to \cite{4657346}.}

	

\textcolor{red}{In Figure. \ref {fig_4}, we plot the sum rate versus  ${K}_{r}/({K}_{r}+K_{t})$. 
For the power-domain IOS, it performs well when $K_r/K_t$ aligns well with the fixed $\epsilon$. In this case, we assume the MUs are symmetrically located and $\tau_{k_r}=\tau_{k_t}$. According to \textit{Proposition 4}, there is no performance gap between the dual-polarized IOS and power-domain IOS when $K_r=K_t$.
However, in some extreme cases, such as when the x-axis coordinate is 0.2 or 0.8, power is wasted due to the mismatch between $\epsilon$ and the number of MUs, resulting in a performance gap of 1.2 bit/s/Hz.  In addition, the dual-polarized RIS-aided system performs well as ${K}_{r}/({K}_{r}+K_{t})$ gradually increases since it can only reflect the signal.}

\textcolor{red}{In Figure. \ref {fig_2}, we present the sum rate versus different XPD factors, considering various numbers of BS antennas and IOS elements. The results show that as the XPD factor between the BS and the IOS, $\beta_{\rm BI}$, increases from 0 to 1, the sum rate first decreases and then increases, which aligns with \textit{Proposition 2}. Furthermore, increasing the number of IOS elements and transmit antennas, compared to the benchmark configuration of $N_t=2, M=4, K=2$, results in improved performance.}

In Figure \ref{fig_3}, the sum rate is plotted versus transmit power for varying XPD factors. As transmit power increases, the sum rate improves due to a higher SINR. Increasing the $\beta_{\rm BI}$ from 0.1 to 0.2 results in performance degradation. This is due to increased polarization leakage, which not only reduces signal power but also increases cross-polarization interference, lowering the SINR. When $\beta_{\rm BI}$ increases to 0.8,  the performance surpasses that of the other cases. This is because the zero elements described in \textit{Proposition 1} increase, leading to more power being allocated to the orthogonal polarization. These results are consistent with the analysis in \textit{Remark 1}.

\section{Conclusion}
In this paper, we considered a dual-polarized IOS-aided
downlink system and investigated the joint BS digital and IOS analog beamforming design to maximize the sum rate for reflective-refractive transmission by utilizing the polarization domain. Then, we analyzed the impact of the XPD and the user ratio on both sides of the dual-polarized IOS. We have the following conclusions: 1) Unlike power-domain IOS, the performance of the dual-polarized IOS is less affected by the ratio of the number of MUs on both sides, as the coupling between refraction and reflection coefficients is decoupled. 2) High polarization leakage may provide extra gains for dual-polarized IOS-assisted systems.
\balance

\bibliographystyle{IEEEtran}
\bibliography{document.bib}
\end{document}